\newcommand{\thisnodetitle}            {Gauge equivalence and  transdimensional perturbation:
                                        point vortices}
\newcommand{\thisnodeauthorgivenname}  {George W.}
\newcommand{\thisnodesauthorsurname}   {Patrick}
\newcommand{\thisnodeaddresslineone}   {Department of Mathematics and Statistics}

\newcommand{\thisnodeaddresslinethree} {University of Saskatchewan}
\newcommand{\thisnodeaddresslinefour}  {Saskatoon, SK S7N 5E6}

\documentclass[fleqn]{article}
\usepackage[totalwidth=6.8in,totalheight=9.25in]{geometry}

\usepackage{amsmath}
\usepackage{amssymb}
\usepackage{amsthm}
\usepackage{amsfonts}
\usepackage{bm}
\usepackage{caption}
\usepackage[mathscr]{euscript}
\usepackage[pdftex]{hyperref}
\usepackage{mathrsfs}
\usepackage{mathtools}\mathtoolsset{showonlyrefs}
\usepackage{titlesec}
\usepackage{xcolor}
\usepackage{enumitem}
\usepackage[noadjust]{cite}
\usepackage{graphicx}
\usepackage[title]{appendix}
\usepackage{comment}
\usepackage{tikz}
\usepackage{tikz-cd}
\usepackage{paralist}

\definecolor{refcolor}{RGB}{42,93,176}
\hypersetup{colorlinks=true,linkcolor=refcolor,citecolor=refcolor}
\newtheorem{theorem}{Theorem}[section]

\usepackage{nodelib}
\newcommand{\leaf}[2]{}
\newcommand{\subnode}[2]{\section{#2}\label{#1}}
\def\[#1\]{\begin{alignat}{16}#1\end{alignat}}
\renewcommand{\nodetextdef}[1]{\emph{#1}}

\newcommand\sing{{\scalerel*{\dag}{X}}}

\excludecomment{notthreadswitch}

\begin{document}

\thispagestyle{empty}
\setlist{nosep}

\begin{raggedright}
{\Large\bf\thisnodetitle}
\\[15pt]{
  \thisnodeauthorgivenname\ \thisnodesauthorsurname\\
  \thisnodeaddresslineone\\
  \thisnodeaddresslinethree\ \thisnodeaddresslinefour\\
  \today
}
\\[10pt]
{\color{lightgray}\rule{\textwidth}{.5pt}}
\parbox{\textwidth}{
  \vspace*{3pt}{\noindent\bf Abstract\\[4pt]}
  \leaf{Abstract}{}
There is an explicit resolution of the Poisson reduction of four planar point vortices, in the
case that three of the vortex strengths are equal and the total vorticity is zero.  The
resolution, a Hamiltonian system on a unified symplectic phase space with a symmetry breaking
parameter, is obtained by appending redundant states. Though single point vortices do not have
the attribute of mass, there are circular assemblages with the collective dynamics of free
massive particles, demonstrating a finite dimensional dynamics where mass emerges from a gauge
symmetry breaking. The internal vibration of these assemblages is coupled to their collective
motion and has the same functional form as the de~Broglie wavelength.

}
\\[4pt]
{\color{lightgray}\rule{\textwidth}{.5pt}}\\[6pt]
\end{raggedright}
\vspace*{1pt}

\section{Introduction}
\leaf{}{}
As is very well known, equilibria of Hamiltonian systems occur at critical points of the energy
function, and their stability may be established by utilizing energy as a Lyapunov function. The
stability is not asymptotic, but rather sufficiently close initial conditions yield arbitrarily
near solutions. At an equilibrium the Hamiltonian vector field is zero and so admits a
linearization. Energy stability implies spectral stability and KAM theory may be applicable in
the converse.

\begin{figure}[h]
\begin{center}\includegraphics{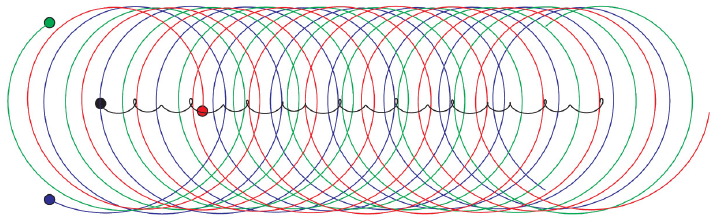}\end{center}
\caption{
  \label{fig:1-graphic-O-alpha-animation}
  Collective motion of $\upO_{\alpha_e}$.}
\end{figure}

\leaf{}{} 
In the case of a continuous symmetry, reductions involve both quotient and momentum, with
purpose to arrive at a generic Hamiltonian system, and again there may be equilibria, (relative
equilibria of the original system where the solution is by a one parameter subgroup of the
symmetry group). Here, though, the symplectic leaves may vary and energy stability does not
directly account for nearby solutions if momentum is altered~\cite{PatrickGW-1992-1}. For
example, with the compact symmetry group~$\SO3$, the generic reduced spaces have dimension~$4$
less, but $6$~less at zero momentum, and the orientation acquires a slow dynamics which can be
modeled as a Lagrangian system~\cite{PatrickGW-1995-1,PatrickGW-1999-1}. If the symmetry is
noncompact then solutions near a singular leaf might escape in nearby leaves (Example~5
of \cite{PatrickGW-RobertsRM-WulffC-2004-1}), and in the physically relevant model of the
Kirchhoff approximation for the dynamics of submerged rigid
objects~\cite{LeonardNE-1997-1,LeonardNE-MarsdenJE-1997-1}, dynamic stability is established
instead by KAM~methods after a \nodetextdef{resolution} of the small symplectic
leaves~\cite{PatrickGW-2003-1,PatrickGW-RobertsRM-WulffC-2008-1}. The planar 4-vortex
system~\cite{PatrickGW-2000-1,PatrickGW-2000-2}, has relative equilibria (here designated
$\upO_{\alpha_e}$) where three vortices of strength $-\Gamma/3$ in the shape of an equilateral
triangle rotate in a circle radius $\alpha_e$ around a central vortex of strength $\Gamma$. The
symmetry is the (noncompact) planar Euclidean group $\SE2$, and the translational momentum $p$
at $\upO_{\alpha_e}$ is zero. At small $p\ne0$, the shape of $\upO_{\alpha_e}$ becomes
vibrational and is observed in simulation~(Fig.~\ref{fig:1-graphic-O-alpha-animation}) to crawl
at velocity $p/m_e$, where $m_e=8\uppi\alpha_e^2/3$. The stability of this motion is an open
problem.

\begin{figure}[h]
\begin{center}\includegraphics{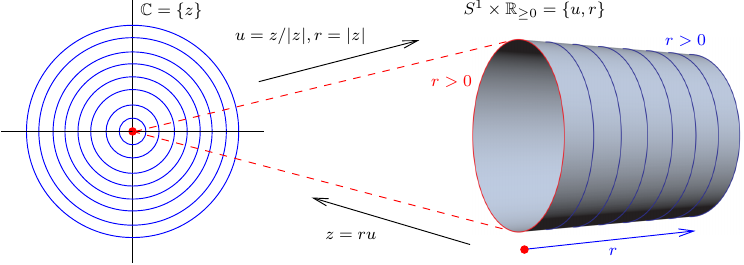}\end{center}
\caption{
  \label{1-graphic-resolution-standard-SO2-action}
  Resolution of the standard $\SO2$ action on $\bbC$.}
\end{figure}

\leaf{}{}
The meaning of \nodetextdef{resolution} may be llustrated by $\SO2=\set{A\in\bbC}{\abs A=1}$
acting by multiplication on $\bbC=\sset{z}$
(Fig.~\ref{1-graphic-resolution-standard-SO2-action}). The \nodetextdef{singularity} at $z=0$
has isotropy $\SO2$, and the complement $z\ne0$ (\nodetextdef{regular sector}) has trivial
isotropies. The resolution consists of the manifold with boundary $\bbR_{\ge0}\times
S^1=\set{(r,u)}{r\ge0,\abs u=1}$ together with the maps $r=\abs z,u=z/\abs z$ and $z=ru$. The
first map is a diffeomorphism to the interior $r>0$, while the second is a surjection
restricting to the inverse of the first. A space of physical states $\bbC=\sset{z}$ is replaced
by one on $\bbR_{\ge0}\times S^1$ via covariance followed by smooth extension from $r>0$ to all
of $r\ge0$, and $z=0$ becomes multiply represented by the circle of states on the boundary
$r=0$. Physical conclusions associated to any two boundary states must agree (\nodetextdef{gauge
equivalence}), a redundancy which can be represented by the obvious action of $\SO2$ on the
boundary (\nodetextdef{gauge symmetry}).

\leaf{}{}
Realizations of Poisson manifolds occur as far back as Lie~(see \cite{WeinsteinA-1983-1}), and
in~\cite[Def.~1.9.1]{DufourJP-ZungNT-2005-1} they are surjective Poisson submersions
$(M,\omega)\to(P,\sset\relax)$ from a symplectic manifold. Here, in the 4-vortex problem, and
denoting the (open dense) regular part of $P$ as $P^\circ$ and the complementary singular part
as $P^\sing$, a resolution is a manifold with boundary $A$, a symplectic manifold $(M,\omega)$,
and a surjective Poisson map $\varphi\colon A\times M\to P$, such that $\varphi(\partial A\times
M)\subseteq P^\sing$ and $\varphi\colon(\onm{int}A)\times M\to P^\circ$ is a Poisson
isomorphism. A dynamics generated on $P$ by a Hamiltonian $H$ is then covered by the dynamics on
$A\times M$ generated by $H\circ\varphi$, and if $\varphi$ is not injective on
$(\onm{int}A)\times M$ then states in $P^\sing$ are multiply represented. If $P^\sing$ is an
embedded submanifold of $P$ then it is naturally a Poisson manifold by function extension and
$\varphi\rto{\partial A\times M}$ is a Poisson submersion ie it is the reduction of $\partial
A\times M$ by the level-set equivalence relation of $\varphi$ in $P^\sing$.

\subnode{2-subnode-3-explicit-resolution}{Explicit resolution}
\leaf{}{}
The planar Euclidean group $\SE2\deq\set{(A,a)}{\mbox{$A\in\bbC$, $\abs A=1$, $a\in\bbC$}}$
with
\[
  \mbox{identity}=(1,0),
\quad
  (A,a)(B,b)\deq(AB,a+Ab),
\quad
  (A,a)^{-1}=(A^{-1},-A^{-1}a),
\]
acts on $\bbC$ (\nodetextdef{standard action}) by $(A,a)z=Az+a$ (the $\SE2$ multiplication is
determined by the requirement that this is a group action). The manifold structure of $\SE2$ is
as a submanifold of $\bbR^4\equiv\bbC^2$. The Lie algebra and corresponding
identifications~\cite{MarsdenJE-RatiuTS-1999-1} are
\[
  &\se2=\bigset{(u,v)}{\mbox{$u\in\bbR$, $v\in\bbR^2$}},
  \qquad
  (u,v)\equiv\frac d{dt}\bigl[5]|_{t=0}(1+\upi ut,tv)\equiv(\upi u,v),
\\
  &\se2^*=\bigset{(\mu,\nu)}{\mbox{$\mu\in\bbR$, $\nu\in\bbC$}},
  \qquad
  \langle{\,(\mu,\nu),(u,v)}\,\rangle\deq\mu u+\nu\cdot v,
\]
the infinitesimal generator and exponential map are
\[
  &\frac d{dt}\bigr[5]|_{t=0}(1+\upi utz+tv)=\upi uz+v,
  \label{eq:se2-C-infinitesimal-generator}
\\
  &\onm{exp}(u,v)=(A,a),
  \qquad
  A=e^{\upi u},
  \qquad
  a=\begin{cases}
  \dfrac{(e^{\upi u}-1)v}{\upi u},&u\ne 0,\\v,&u=0,
  \end{cases}
\]
and the adjoint and coadjoint actions are
\[
  &\onm{Ad}_{(A,a)}(u',v')=\bigl(u',Av'-\upi u' a\bigr),
  &&\qquad
  \onm{ad}_{(u,v)}(u',v')=\bigl(0,\upi (uv'-u'v)\bigr),
\\
  &\onm{CoAd}_{(A,a)}(\mu,\nu)=(\mu+a\wedge A\nu,A\nu),
  &&\qquad
  \onm{coad}_{(u,v)}(\mu,\nu)=(v\wedge\nu,\upi u\nu),
\]
where $z\wedge w\deq-\onm{Im}(z\bar w)$ (and also there is the notation $z,w\in\bbC$ $z\cdot
w\deq\onm{Re}(z\bar w)$). The coadjoint isotropy groups are
\[
  \SE2_{(\mu,\nu)}=
  \begin{cases}
    \set{A=1,a=t\nu}{t\in\bbR},&\nu\ne0,
  \\[2pt]
    \SE2,&\nu=0,
  \end{cases}
\]
and these are two $\SE2$ conjugacy classes, so there are two isotypic subsets of momenta
$\se2^{*\circ}=\set{\mu,\nu)}{\nu\ne0}$ (\nodetextdef{generic}, or \nodetextdef{regular}) and
$\se2^{*\sing}=\set{\mu,\nu)}{\nu=0}$ (\nodetextdef{rotational} or \nodetextdef{singular}).

\begin{figure}[h]\begin{center}
\includegraphics{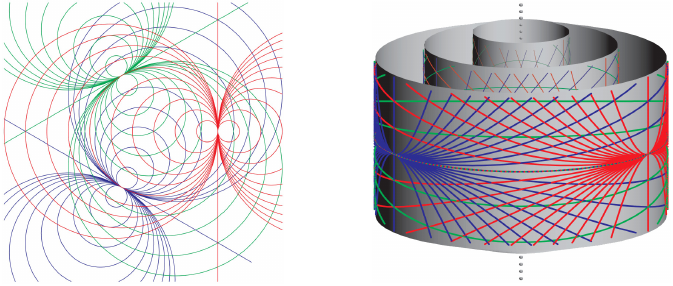}
\caption{\label{fig:1-graphic-SE2-adjoint-coadjoint-action}Standard and adjoint actions.}
\end{center}\end{figure}

\leaf{}{}
The flow lines of the exponential of $\SE2$ through the standard action
(Fig.~\ref{fig:1-graphic-SE2-adjoint-coadjoint-action} left) are rotations of $z$, radius
$r=\abs{z-\upi v/u}$, with center of rotation at $\upi v/u$, and angular frequency $u$, or if
$u=0$ then translations at velocity $v$. For constant $v$ and $u\to0$, the circle centers limit
to infinity in the direction $\upi v$, the velocity at $z$ to $v$, and the motion to
straight-line with velocity $v$, thus smoothly interpolating plane rotation and translation. An
element $(u,v)\in\se2$ may be thought of as a mixture of rotation and translation, with $u=0$
corresponding to pure translation. The adjoint action is a similar picture in each constant $u'$
plane, except in $u'=0$ (pure translation) there are only rotations about the origin (pure
translational velocity has simple rotational covariance). The flow lines of the coadjoint action
(Fig.~\ref{fig:1-graphic-SE2-adjoint-coadjoint-action} right) are intersections of planes with
the constant~$\abs\nu$ cylinders and $\nu=0$ is an axis of fixed points.  If $\SE2$ is brought
to act on a physical state corresponding to one parameter subgroup of $\SE2$, then the generator
changes similarly to the standard action (above left) while the momentum changes in an entirely
different manner eg translation can unboundedly increase moment arms. This does not occur for
compact symmetry groups because their Lie algebras have an invariant metric and the two actions
are isomorphic (and the adjoint and coadjoint orbits are compact). As is well known, in the case
of noncompact symmetry, generator and momentum covariance are distinct.

\leaf{}{}
Given an index set $\clN$, and nonzero $\Gamma_n$, $n\in\clN$, the \nodetextdef{plane vortex
system}~\cite{NewtonPK-2001-1} is a Hamiltonian system with phase space
\[
  &P\deq\set{z\in\bbC^\clN}{\mbox{$z_n\ne z_m\;\forall m,n$}},
\\
  &z_n=x_n+\upi y_n,
\\
  &\omega\deq\sum_n\Gamma_n\,dx_n\wedge dy_n=\frac\upi2\sum_n\Gamma_n\,dz_n\wedge d\bar z_n,
\\
  &\{f,g\}=\sum_n\frac1{\Gamma_n}\bigl[4](\frac{\pl f}{\pl x_n}\frac{\pl g}{\pl y_n}
    -\frac{\pl f}{\pl y_n}\frac{\pl g}{\pl x_n}\bigr[4])
  = \sum_n\frac1{\Gamma_n}\{f,g\}_{z_n},
\\
  &H\deq-\frac1{8\uppi}\sum_{m\ne n}\Gamma_m\Gamma_n\onm{ln}\abs{z_m-z_n}^2,
\\
  &\frac{dz_n}{dt}
  =-\frac{2\upi}{\Gamma_n}\frac{\pl H}{\pl\bar z_n}
  =\frac\upi{2\uppi}\sum_{n\ne m}\frac{\Gamma_m(z_n-z_m)}{\abs{z_n-z_m}^2},
  \label{eq:plane-vortex-hamiltonian-system}
\]
(see Appendix~\ref{subnode:wirtinger-derivatives} for a summary of Wirtinger derivatives as they
are used here).  The system is symmetric under the diagonal action of $\SE2$, and a momentum
mapping is
\[
  \label{eq:z-momentum}
  J\colon P\to\se2^*,
  \quad
  \mu=-\frac12\sum_n\Gamma_n\abs{z_n}^2,
  \quad
  \nu=-\upi\,\sum_n\Gamma_nz_n.
\]
$J$ is equivariant with respect to the action
$(\mu,\nu)\mapsto\onm{CoAd}_{(A,a)}(\mu,\nu)+\sigma(A,a)$, where the \nodetextdef{cocycle}
$\sigma\colon\SE2\to\se2^*$ is \cite[Def.~4.2.4]{AbrahamR-MarsdenJE-1978-1}
\[
  \sigma(A,a)&=\bigl(J((Az_n+a)_n)-\onm{CoAd}_{(A,a)}J(z)\bigr)\bigr[3]|_{z=0}
  &=-\bigl[4](\sum_n\Gamma_n\bigr[4])\bigl[4](\frac12|a|^2,ia\bigr[4]).
\]

\leaf{}{}
Assume $n$ from $0,\ldots,N$, $\Gamma\ne 0$, $\Gamma_0=\Gamma$, and $\Gamma_n=-\Gamma/N$, $n\ge
1$, so that $\sum_n\Gamma_n=0$ and the momentum is $\onm{CoAd}$-equivariant. The
symmetric group $S_n$ acts symplectically by permutation of $z_1,\ldots,z_n$. Define the $\SE2$
and flow invariant submanifolds $P^\sing\deq\set{z\in P}{\nu=0}$ (\nodetextdef{rotational
sector}) and $P^\circ\deq P\setminus P^\sing$ (\nodetextdef{regular sector}), note that
\[
  \nu=-\upi\Gamma z_0+\sum_{n=1}\frac{\upi\Gamma}N\,z_n
  =\frac{\upi\Gamma}N\sum_{n=1}(z_n-z_0)=0
  \quad\Leftrightarrow\quad\sum_{n=1}(z_n-z_0)=0,
\]
and define
\[
  &P_1\deq\bigset{u=(u_0,u_1,\ldots,u_N)}{\mbox{$u_0\ge0$ and $\sum_{n=1} u_n=0$,
   $u_m\ne u_n\;\forall m,n\ge 1$}},
\\
  &P_1^\sing=\bigset{u}{u_0=0},
\\
  &P_1^\circ\deq P_1\setminus P_1^\sing.
\]
$P_1$ is 
a manifold with boundary $P_1^\sing$ and interior $P_1^\circ$.  Define the $\SE2$
invariant map $\pi_{P^\circ}\colon P^\circ\to P_1^\circ$ by
\[
  \label{eq:z-to-u-reduction}
  \pi_{P^\circ}:
  \quad
  u_0=\bigabs[4]{\mathinner{\sum_{m=1}(z_m-z_0)}},
  \quad
  u_n=u_0(z_n-z_0)\bigl[6](\mathinner{\sum_{m=1}(z_m-z_0)}\bigr[6])^{\!-1}-\frac{u_0}N.
\]
The map $\pi_{P^\circ}$ translates $z_0$ to the origin, the division removes rotation, and
$P^\circ$ and $P^\sing$ are mapped into $P_1^\circ$ and $P_1^\sing$ respectively.

\leaf{}{}
To find a right inverse of $\pi_{P^\circ}$, assume $u\in P_1$, choose $z_0=0$, and obtain
from the second of~\eqref{eq:z-to-u-reduction} the eigenvalue problem
\[
  \bigl[4](\mathinner{u_n+\frac{u_0}N}\bigr[4])\sum_{m=1}z_m=u_0z_n.
\]
The matrix corresponding to the left has the $N-1$ dimensional nullspace $z_1+\cdots+z_N=0$, and
there is one $u_0$ eigenvector, namely $z_n=u_n+u_0/N$ because
\[
  \bigl[4](\mathinner{u_n+\frac{u_0}N}\bigr[4])\sum_{m=1}\bigl[4](u_m+\frac{u_0}N\bigr[3])
  =\bigl[4](u_n+\frac{u_0}N\bigr[4])u_0,
\]
so $z_n=a(u_n+u_0/N)$ where $a$ is an undetermined complex number. Substituting into the first
of~\eqref{eq:z-to-u-reduction} gives
\[
  u_0=\bigabs[5]{\sum_{n=1}a\bigl[4](u_n+\frac{u_0}N\bigr[4])}
  =u_0\mathinner{\abs{a}},
\]
so $\abs a=1$, and choosing $a=1$ obtains
\[
  \label{eq:u-section}
  \iota_{P_1}\colon P_1\to P,\qquad z_0=0,\qquad z_n=u_n+\frac{u_0}N.
\]
Although $\pi_{P^\circ}$ is not defined on the rotational sector $P^\sing$, the map
$\iota_{P_1}$ is smooth on all of $P_1$. The restriction $\iota_{P_1}\colon P_1^\circ\to
P^\circ$ is a right inverse for $\pi_{P^\circ}$ and a global section to the $\SE2$ action on
$P^0$ ie for $u_0>0$ the image of the right inverse is an embedded submanifold meeting every
regular sector $\SE2$ orbit exactly once and $\pi_{P^\circ}\colon P^\circ\to P_1^\circ$ is a
trivial left principle bundle. The map $\iota_{P_1}$ is not unique: if $g(u)$ is any smooth
$\SE2$-valued function then $u\mapsto g(u)\,\iota_{P_1}(u)$ is another. Following general
practice, such a map is a \nodetextdef{gauge}, \eqref{eq:u-section} will be called
the \nodetextdef{standard gauge}, a $g(u)$ is a \nodetextdef{local gauge transformation}, and
a \nodetextdef{gauge transformation} is the special case of where $g(u)$ is constant.
Evolutions $c(t)$ on $P_1^\circ$ may
be \nodetextdef{reconstructed}~\cite[\textsection4.3]{AbrahamR-MarsdenJE-1978-1} to evolutions
on $P^\circ$ as $g(t)\,\iota_{P_1}(c(t))$ where $g(t)$ is a curve in $\SE2$, and reconstruction
does not depend on gauge (the dynamics defined by \eqref{eq:plane-vortex-hamiltonian-system}
does not).

\leaf{}{}
A dynamics on all of $P_1$ will be extablished by imposing that $\pi_{P^\circ}\colon P^\circ\to
P_1^\circ$ is Poisson and, descending the
Hamiltonian \eqref{eq:plane-vortex-hamiltonian-system}, and then extending the bracket and
Hamiltonian to all of $P_1$ by continuity. The Poisson bracket of
\[
  f(u_0,\ldots, u_N,\bar u_0,\ldots,\bar u_N),
  \qquad
  g(u_0,\ldots, u_N,\bar u_0,\ldots,\bar u_N),
\]
is obtained by composition
\[
  \{f,g\}&=-2\upi\sum_{p=0}\frac1{\Gamma_p}\{f,g\}_{z_p}
\\
  &=-2\upi\sum_{p=0}\sum_{m,n}
  \frac1{\Gamma_p}\bigl[5](
  \{u_m,u_n\}_{z_p}\frac{\partial f}{\partial u_m}\,\frac{\partial g}{\partial u_n}
  +\{u_m,\bar u_n\}_{z_p}\frac{\partial f}{\partial u_m}\,
    \frac{\partial g}{\partial\bar u_n}
\\[-4pt]
  &\qquad\qquad\qquad\qquad\mbox{}
  +\{\bar u_m,u_n\}_{z_p}\frac{\partial f}{\partial\bar u_m}\,
    \frac{\partial g}{\partial u_n}
  +\{\bar u_m,\bar u_n\}_{z_p}\frac{\partial f}{\partial\bar u_m}\,
    \frac{\partial g}{\partial\bar u_n}
    \bigr[5])
\\[2pt]
  &=\sum_{m,n}
  \{u_m,u_n\}\frac{\partial f}{\partial u_m}\,\frac{\partial g}{\partial u_n}
  +\{u_m,\bar u_n\}\frac{\partial f}{\partial u_m}\,
    \frac{\partial g}{\partial\bar u_n}
  +\{\bar u_m,u_n\}\frac{\partial f}{\partial\bar u_m}\,
    \frac{\partial g}{\partial u_n}
  +\{\bar u_m,\bar u_n\}\frac{\partial f}{\partial\bar u_m}\,
    \frac{\partial g}{\partial\bar u_n}.
\]
Since
\[
  \sum_{m=1}(z_m-z_0)
  =-Nz_0+\sum_{m=1}z_m
  =-\frac N\Gamma\bigl[5](\Gamma z_0-\sum_{m=1}\frac\Gamma Nz_m\bigr[5])
  =\frac{-\upi N\nu}{\Gamma},
\]
\eqref{eq:z-to-u-reduction} may be written
\[
  \label{eq:z-to-u-reduction-alt}
  u_0=\frac{N\abs\nu}{\abs\Gamma},
  \qquad
  u_n=\frac{u_0}{N\nu}\bigl(\upi\Gamma(z_n-z_0)-\nu\bigr).
\]
The computation is simplified by the expressions \eqref{eq:z-to-u-reduction-alt} because the
Poisson bracket is zero for any function obtained by composition with \eqref{eq:z-momentum} (ie
a function $\se2^*=\sset{(\mu,\nu)}$) and any $\SE2$ invariant function. In particular,
$\{u_m,u_0\}=0$, $m=0,\ldots,N$, and for $m,n\ne0$
\[
  \{u_m,u_n\}
  =-2\upi\sum_{p=0}\frac1{\Gamma_p}\,\frac{-u_0^2\Gamma^2}{N^2\nu^2}\,
  \{{z_m-z_0,z_n-z_0}\}_{z_p}=0,
\]
and similarly $\{\bar u_m,\bar u_n\}=0$, while
\[
  &\{u_m,\bar u_n\}
  =-2\upi\sum_{p=0}\frac1{\Gamma_p}\,\frac{-u_0^2\Gamma^2}{N^2\abs\nu^2}
    \,\{{z_m-z_0,\bar z_n-\bar z_0}\}_{z_p}
  =2\upi\bigl[5](\frac1{\Gamma}+\frac1{\Gamma_n}\,\delta_{nm}\bigr[5])
\]
where $\delta_{nm}$ is the Kronecker delta, and also similarly $\{\bar u_m,u_n\}=-\{u_m,\bar
u_n\}$, giving the reduced Poisson bracket
\[
  \label{eq:u-bracket}
  \{f,g\}=-\sum_{p=1}\frac N{\Gamma}\{f,g\}_{u_p}
  +\frac1{\Gamma}\sum_{n,m}
  \bigl[5](
  \frac{\partial f}{\partial u_n}\,
  \frac{\partial g}{\partial\bar u_m}
  -\frac{\partial f}{\partial\bar u_n}\,
  \frac{\partial g}{\partial u_m}
  \bigr[5]).
\]
This must be regarded as a bracket on functions on the reduced space $u_1+\cdots u_N=0$, which
are in correspondence with functions which satisfy
\[
  \label{eq:u-functions}
  f(u_1+a,\ldots,u_N+a)=f(u_1,\ldots,u_N)
\]
ie are invariant under the diagonal addition action of $\bbC$ on $\bbC^N$.  But for such
invariant functions the second term in \eqref{eq:u-bracket} is zero because
\[
  &\sum_n\frac{\pl f}{\pl u_n}
  =\frac d{da}\bigr[5]|_{a=0}f(u_1+a,\ldots,u_N+a)
  =\frac d{da}\bigl[5]|_{a=0}f(u_1,\ldots,u_N)
  =0,
\\
  &\sum_{n,m}\bigl[4](
  \frac{\partial f}{\partial u_n}\,
  \frac{\partial g}{\partial\bar u_m}
  -\frac{\partial f}{\partial\bar u_n}\,
  \frac{\partial g}{\partial u_m}\bigr[4])
  =
  \sum_n\frac{\partial f}{\partial u_n}\times
  \sum_m\frac{\partial f}{\partial\bar u_m}
  -
  \sum_n\frac{\partial f}{\partial\bar u_n}\times
  \sum_m\frac{\partial f}{\partial u_m}
  =0,
\]
so, under the identification, \eqref{eq:u-bracket} is the same bracket as the system of
$N-1$ vortices with strength $\Gamma_p=-\Gamma/N$. Use of the simpler first part
of \eqref{eq:u-bracket} (as is done below) is valid for the bracket of functions on the
reduced Poisson space $u_1+\cdots u_N=0$ that are extended so as to
satisfy \eqref{eq:u-functions}.

\leaf{}{}
Since $u_0$ does not evolve it can be regarded as a parameter in a $6$ dimensional symplectic
space, with cannonical symplectic form if $\Gamma>0$ and the negative of that if $\Gamma>0$. Even
though the quotient map is undefined on the rotational sector, the symplectic form and
Hamiltonian have well defined continuous limits at $u_0=0$.

\leaf{}{}
The Hamiltonian may be pulled back to the reduced space by composition
with \eqref{eq:u-section}, and the sum splits over $m=0$ (in which $z_m=0$) and $m\ge 1$, the
multiplicative constants within the logarithm may be discarded, and rescaling $u\deq u_0/N$,
with result
\[
  \label{eq:u-hamiltonian}
  H&=-\frac1{4\uppi}\sum_{1\le m<n}\Gamma_m\Gamma_n\onm{ln}\abs{u_m-u_n}^2
  -\frac1{4\uppi}\sum_{m=1}\Gamma\,\Gamma_n\onm{ln}\abs{u_m+u}^2
\\
  &=-\frac1{4\uppi}\sum_{1\le m<n}\bigl[4](\frac{\Gamma}{N}\bigr[4])^2
  \onm{ln}\abs{u_m-u_n}^2
  +\frac{\Gamma^2}{4\uppi N}\sum_{m=1}\onm{ln}\abs{u_m+u}^2
\]
ie the energy of the $N$ vortex system (but again the phase space is constrained) plus an $\SO2$
symmetry breaking perturbation.

\leaf{}{}
By way of summary, the above may be regarded as a form of Poisson reduction of the point vortex
system, where the regular part of the Poisson manifold is realized, while the singular part is
the boundary of that and retains a symmetry. To compare with existing well-developed Poisson
reduction theory, see~\cite{DufourJP-ZungNT-2005-1,MarsdenJE-RatiuTS-1986-1, WeinsteinA-1987-1}.

\begin{theorem}
\label{thm:resolution-summary}
The phase space $P$ is the disjoint union of the $\SE2$ invariant Poisson submanifolds $P^\circ$
and $P^\sing$, and the phase space $P_1$ is the disjoint union of $P_1^\circ$ and $P_1^\sing$.
The map $\pi_{P^\circ}\colon P^\circ\to P_1^\circ$ defined by \eqref{eq:z-to-u-reduction} is a
Poisson quotient for the action of $\SE2$ on $P^\circ$ ie $\pi_{P^\circ}$ is a Poisson
submersion such that $\pi_{P^\circ}(z)=\pi_{P^\circ}(z')$ if and only if there is a
$(A,a)\in\SE2$ such that $(A,a)z=z'$.  The map $\iota_P\colon P_1\to P$ is an injective
immersion which respects the decompositions $P=P^\sing\cup P^\circ$ and $P_1=P_1^\sing\cup
P_1^\circ$, the $\SE2$ stabilizer of $P^\sing$ is the natural copy of $\SO2$ in $\SE2$,
$\iota_P$ is $\SO2$ intertwining on $P_1^\circ$, and induces a diffeomorphism
$P_1^\sing/\SO2\simeq P^\sing/\SE2$. Moreover,
\smallskip
\begin{compactenum}
\item
for all integral curves $c(t)\in P_1^\sing$, $t\in(a,b)$, there is a curve $g(t)\in\SE2$ such
that $g(t)\,\iota_{P_1}(c(t))$ is an integral curve in $P^\sing$; and
\item
for all integral curves $d(t)\in P^\sing$, $t\in(a,b)$, there is an integral curve $c(t)\in
P_1^\sing$ and a curve $g(t)\in\SE2$ such that $d(t)=g(t)\,\iota_{P_1}\circ c(t)$.
\end{compactenum}
\end{theorem}

\leaf*{Proof}{}  
For~(1), let $t_0\in(a,b)$, $p=c(t_0)$, $\hat c(t)\deq\iota_{P_1}\circ c$, and $\hat p\deq\hat
c(t_0)$. Choose a sequence $\seq{p_i\in P_1^\circ}$ converging to $p$ and let $\hat
p_i\deq\iota_{P_1}(p_i)$. Since the flow on $P_1$ has open domain and is continuous, for large
enough $i$ there are integral curves $c_i\colon(a,b)\to P_1^\circ$ such that $c_i(t_0)=p_i$, and
$c_i\to c$ pointwise. Since $\pi_{P^\circ}$ is Poisson, there are integral curves $\hat
d_i\colon(a,b)\to P$ such that $\hat d_i(t_0)=\hat p_i$ and $\pi_{P^\circ}\circ d_i=c_i$, and
then there is an integral curve $d\colon(a,b)\to P^\sing$ such that $d(t_0)=\hat p$.
$\pi_{P^\circ}\circ d_i=\pi_{P^\circ}\circ\hat c_i$, so for any particular $t$ there is a
sequence $\seq{g_i}$ such that $d_i(t)=g_i\hat c_i(t)$. $\SE2$ acts freely and properly, so some
subsequence of $g_i$ converges, hence $d(t)$ and $\hat c(t)$ are in the same $\SE2$ orbit, and
there is a unique curve $g(t)$ such that $g(t)\hat c(t)=d(t)$. The converse~(2) is
similar.~\nodeqed

\leaf{}{}
For $N=3$ the reduced phase space may be cannonicalized using
\[\label{eq:u-to-v-dim3-reduction}
  u_1=v_1+v_2,
  \qquad
  u_2=\uptheta v_1+\uptheta^2 v_2,
  \qquad
  u_3=\uptheta^2 v_1+\uptheta v_2,
  \qquad
  \uptheta\deq e^{2\uppi\upi/3},
\]
which is from $\bbC^2=\sset{(v_1,v_2)}$ into $u_1+u_2+u_3=0$ because the third primitve root of
unity satisfies
\[
  1+\uptheta+\uptheta^2=\frac{1-\uptheta^3}{1-\uptheta}=0.
\]
Altering \eqref{eq:u-to-v-dim3-reduction} to the equations
\[
  u_1=v_0+v_1+v_2,
  \qquad
  u_2=v_0+\uptheta v_1+\uptheta^2 v_2,
  \qquad
  u_3=v_0+\uptheta^2 v_1+\uptheta v_2,
\]
creates an invertible linear operator on $\bbC^3$ because $v_0$ moves orthogonally off the
reduced space along its normal $(1,1,1)$. The inverse is therefore a diffeomorphism from
$u_1+u_2+u_3=0$ to $v_0=0$ and it corresponds to the lowerhand $2\times 3$ matrix of
\[
  \bmat{1&1&1\\1&\uptheta&\uptheta^2\\1&\uptheta^2&\uptheta}^{-1}
  =\frac13\bmat{1&1&1\\1&\uptheta^2&\uptheta\\1&\uptheta&\uptheta^2},
\]
and so
\[
  v_1=\frac13(u_1+\uptheta^2u_2+\uptheta u_3),
  \qquad
  v_2=\frac13(u_1+\uptheta u_2+\uptheta^2u_3),
\]
together with \eqref{eq:u-to-v-dim3-reduction} identifies the reduced space as
$\bbC^2=\sset{v_1,v_2}$. The Poisson bracket of $f(v_1,v_2)$ and $g(v_1,v_2)$ ($v_1$ and $v_2$
satisfy \eqref{eq:u-functions})
\[
  \lbrace v_1,\bar v_1\rbrace&=
  \frac19\,
  \lbrace{u_1+\uptheta^2u_2+\uptheta u_3,
  \bar u_1+\bar\uptheta^2u_2+\bar\uptheta u_3\rbrace}
\\
  &=\frac19\bigl(
  \lbrace u_1,\bar u_1\rbrace+\lbrace u_2,\bar u_2\rbrace
  +\lbrace u_3,\bar u_3\rbrace\bigr)
\\
  &=\frac19\times\frac{-3}\Gamma\times 3
\\
  &=-\frac1\Gamma
\]
\vspace{-\belowdisplayskip}\vspace{-\abovedisplayskip}
\[
  \lbrace v_1,\bar v_2\rbrace&=
  \frac19\,
  \lbrace{u_1+\uptheta^2u_2+\uptheta u_3,
  \bar u_1+\bar\uptheta u_2+\bar\uptheta^2u_3\rbrace}
\\
  &=\frac19\bigl(
  \lbrace u_1,\bar u_1\rbrace+\uptheta\lbrace u_2,\bar u_2\rbrace
  +\uptheta^2\lbrace u_3,\bar u_3\rbrace\bigr)
\\
  &=\frac19\times\frac{-3}\Gamma\times 0
\\
  &=0,
\]
and similarly $\lbrace v_2,\bar v_2\rbrace=-1/\Gamma$ and $\lbrace v_2,\bar v_1\rbrace=0$, so
\[
  &\{f,g\}=
  -\frac1\Gamma\{f,g\}_{v_1}-\frac1\Gamma\{f,g\}_{v_2}.
\]
The Hamiltonian is by \eqref{eq:u-hamiltonian} and
\eqref{eq:u-to-v-dim3-reduction} (up to a constant)
\[
  &(u_1-u_2)(u_1-u_3)(u_2-u_3)=3\sqrt3(v_1^3-v_2^3)\,\upi,
\\
  &(u_1+u)(u_2+u)(u_3+u)=v_1^3+v_2^3-3uv_1v_2+u^3,
\]
\vspace{-\abovedisplayskip}\vspace{-\belowdisplayskip}
\[
  H&=-\frac1{4\uppi}\sum_{1\le m<n}\bigl[4](\frac{\Gamma}{N}\bigr[4])^2
  \onm{ln}\abs{u_m-u_n}^2
  +\frac{\Gamma^2}{4\uppi N}\sum_{m=1}\onm{ln}\abs{u_m+u}^2
\\
  &=-\frac{\Gamma^2}{36\uppi}
  \onm{ln}\bigl(\abs{(u_1-u_2)(u_1-u_3)(u_2-u_3)}^2\bigr)
  +\frac{\Gamma^2}{12\uppi}\onm{ln}\bigl(\abs{(u_1+u)(u_2+u)(u_3+u)}^2\bigr)
\\
  &=-\frac{\Gamma^2}{36\uppi}\bigl[4](
  \onm{ln}\bigl(\bigabs[1]{v_1^3-v_2^3}^2\bigr)
  -3\onm{ln}\bigl(\bigabs[1]{v_1^3+v_2^3-3uv_1v_2+u^3}^2\bigr)\bigr[4])
\\
  \label{eq:v-hamiltonian}
  &=-\frac{\Gamma^2}{18\uppi}\bigl[4](
  \onm{ln}\bigabs[4]{\frac{v_1^3-v_2^3}{(v_1^3+v_2^3)^3}}
  -3\onm{ln}\bigabs[4]{1-\frac{(3v_1v_2-u^2)u}{v_1^3+v_2^3}}\bigr[4]).
\]
The permutation group $S_3$ acts on the space $P^1$ by fixing $u_0$ and permuting
$(u_1,u_2,u_3)$, and both $\uppi^0$ and $i^0$ intertwine with the the action of $S^3$ on
$P^0$. From \eqref{eq:u-to-v-dim3-reduction} the action of $S_3$ on $(v_1,v_2)$, determined by
the generators $(1,2,3)$ and $(2,3)$ is
\[
  \label{eq:S1-v-action}
  (1,2,3)(v_1,v_2)=(\uptheta v_1,\uptheta^{-1}v_2),
  \qquad
  (2,3)(v_1,v_2)=(v_2,v_1).
\]

\begin{notthreadswitch}

\leaf{}{}
As will be shown, the pair $\uppi^0$,
$\iota^0$ \nodetextdef{regularizes} the generic sector, in that
\smallskip
\begin{compactenum}
\item
the reduction $\uppi^0\colon\clI P^0\to\clI P^1$ is to the product of a constant symplectic
manifold and the set of dynamical constants $\bbR_{>0}=\sset{u_0>0}$; and
\item
the symplectic form and Hamiltonian at $u_0=0$ forms a $\SO2$ symmetric \nodetextdef{limit
system} on $\partial P^1$, the dynamics of which corresponds to the dynamics of the original
system on the rotational sector (the additional \nodetextdef{gauge symmetry} of the limit system
corresponds to gauge transformation); and
\item
the Hamiltonian reductions of the limit system by its gauge symmetry correspond to those of the
original system with $\nu=0$.
\end{compactenum}
\smallskip
In this way, the small $\nu$ generic sector will be realized as a symmetry breaking perturbation
of the more symmetric rotational sector. The momentum of the standard gauge is
\[
  &\mu=-\frac12\sum_{n=1}^N\Gamma_n\bigabs[4]{u_n+\frac{u_0}N}^2
  =\frac\Gamma{2N}\sum_{n=1}\abs{u_n}^2+\frac{\Gamma u_0^2}{2N^2},
\\
  &\nu=-i\sum_{n=1}\Gamma_n\bigl[3](u_n+\frac{u_0}N\bigr[3])=\frac{\upi\Gamma u_0}N
\]
so for $\nu>0$ and $u_0=N\nu/\abs\Gamma$ the standard gauge maps into the $\nu$-momentum level
set. Other momenta can be obtained by gauge transformation.

\leaf{}{}
The decreasing chain of submanifolds $P^0\supseteq\partial P^0\supseteq\emptyset$ is
a \nodetextdef{stratification} ie a decreasing chain of submanifolds such that the sequential
set differences are submanifolds, the connected components of which are the \nodetextdef{strata}
cf~\cite[Def.~1.1.1]{TrotmanD-2020-1}. Here there are the two generic and rotational strata
\[
  P^0\setminus\partial P^0=\clI P^0,
  \qquad
  \partial P^0\setminus\emptyset=\partial P^0,
\]
and analogously for the stratification $P^1\supseteq\partial P^1\supseteq\emptyset$. The
regularization can be summarized by
$$
\begin{tikzcd}
  \partial P^0\arrow[hook,r]
  &P^0
  &\clI P^0
  \arrow[hook,l]
  \arrow[rightarrow,dr,"\uppi^0"]
\\
  \partial P^1
  \arrow[hook,r]\arrow[u,"\iota^1" left,"\rotatebox{90}{$\sim$}" right]
  &P^1
  \arrow[u,"\iota^1"]
  &\clI P^1
  \arrow[hook,l]
  \arrow[u,"\iota^1"]
  \arrow[u,"\iota^1"]
  &\clI P^1
  \arrow[equal,l]
\end{tikzcd}
$$
where at the right $\partial P^0$ and $\partial P^1$ have corresponding symplectic reductions.
For the case $\SE2$ the limit system $\partial P^1$ is (parameterized) Hamiltonian with symmetry
the abelian $\SO2$ for which there is only one momentum type, but for more complicated groups
such as $\SE3$ there could be a further regularization of $\partial P^1$ to $\partial P^2$.

\leaf{}{}
Generalizing, let the permulation group $S_N$ act on $\bbC^N$ by
\[
  \sigma\,(v_1,\ldots,v_N)=(v_{\sigma^{-1}(1)},\ldots,v_{\sigma^{-1}(N)})
\]
and let $\sigma$ be the permutation that fixes $1$ but otherwise shifts
right ie
\[
  \sigma=\pmat{1&2&3&4&\cdots&N-1&N\\1&3&4&5&\cdots&N&2},
\]
let $\uptheta=e^{2\uppi i/N}$, and let
\[
  \bm\uptheta=\pmat{1,\uptheta,\uptheta^2,\cdots,\uptheta^{N-1}},
\]
regarded as a column vector, so \eqref{eq:u-to-v-dim3-reduction} generalizes as
\[
  \label{eq:u-to-v-reduction}
   u_i
  =v_1\,\sigma^0\bm\uptheta
  +v_2\,\sigma^1\bm\uptheta
  +\cdots
  +v_N\,\sigma^{N-1}\bm\uptheta
\]
and the corresponding matrix (left) and inverse (right) is
\[
  \bmat{
  1      & 1            & 1             & \cdots & 1        \\
  1      & \uptheta       & \uptheta^{N-1}  & \cdots & \uptheta^2 \\
  1      & \uptheta^2     & \uptheta        & \cdots & \uptheta^3 \\
  \vdots & \vdots       & \vdots&\cdots & \vdots            \\
  1      & \uptheta^{N-1} & \uptheta^{N-2}  & \cdots &\uptheta
  }^{-1}
  =\;
 \frac1N\bmat{
  1      & 1            & 1             & \cdots & 1        \\
  1      & 1+Na         & 1             & \cdots & 1+Nb     \\
  1      & 1+Nb         & 1+Na          & \cdots & 1        \\
  \vdots & \vdots       & \vdots&\cdots & \vdots            \\
  1      & 1            & 1             & \cdots&  1+Na
  },
\quad
  a=-\frac1{1-\uptheta},
\quad
  b=\frac{\uptheta}{1-\uptheta}.
\]
Splitting the inverse as a sum of the matrix of all $1$'s (which copies of the average
$\mathring v$) and a matrix involving $a$ and $b$, and defining
\[
  \bm\nu=\pmat{0&a&b&0&\ldots&0},
\]
also regarded as a column vector, the inverse of \eqref{eq:u-to-v-reduction} is
\[
  v_i
  =\mathring v
  +u_1\,\sigma^0\bm\nu
  +u_2\,\sigma^1\bm\nu
  +\cdots
  +u_N\,\sigma^{N-1}\bm\nu.
\]

\end{notthreadswitch}

\leaf{}{}
If $u=0$ then \eqref{eq:v-hamiltonian} is
\[
  \label{eq:v-u0-hamiltonian}
  H=-\frac{\Gamma^2}{18\uppi}\onm{ln}\bigabs[4]{\frac{v_1^3-v_2^3}{(v_1^3+v_2^3)^3}}
\]
and this is invariant under the diagonal action of $\SO2$ ie
$e^{\upi\uptheta}(v_1,v_2)=(e^{\upi\uptheta}v_1,e^{\upi\uptheta}v_2)$ because
\[
  \bigabs[7]{\frac{(e^{\upi\uptheta}v_1)^3-(e^{\upi\uptheta}v_2)^3}
  {\bigl((e^{\upi\uptheta}v_1)^3+(e^{\upi\uptheta}v_2)^3\bigr)^3}}
  =\bigabs[7]{e^{6\upi\uptheta}\frac{v_1^3-v_2^3}{(v_1^3+v_2^3)^3}}
  =\bigabs[7]{\frac{v_1^3-v_2^3}{(v_1^3+v_2^3)^3}}.
\]
The Hamiltonian vector field of \eqref{eq:v-u0-hamiltonian} is
\[
  &\frac{dv_1}{dt}
  =\frac{-1}\Gamma\times-2\upi\,\frac{\pl H}{\pl\bar v_1}
  =\frac{2\upi}\Gamma\frac{\pl H}{\pl\bar v_1}
\\
  &H=-\frac{\Gamma^2}{36\uppi}\onm{ln}\bigabs[4]{\frac{v_1^3-v_2^3}{(v_1^3+v_2^3)^3}}^2,
\]
\vspace{-\abovedisplayskip}\vspace{-\belowdisplayskip}
\[
  \frac{\pl H}{\pl\bar v_1}&=
  -\frac{\Gamma^2}{36\uppi}\,
  \bigabs[4]{\frac{(v_1^3+v_2^3)^3}{v_1^3-v_2^3}}^2\,
  \frac{v_1^3-v_2^3}{(v_1^3+v_2^3)^3}\,
  \frac\pl{\pl \bar v_1}\bigl[4](\frac{\bar v_1^3-\bar v_2^3}{(\bar v_1^3+\bar v_2^3)^3}\bigr[4])
\\
  &=
  -\frac{\Gamma^2}{36\uppi}\,
  \bigabs[4]{\frac{(v_1^3+v_2^3)^3}{v_1^3-v_2^3}}^2\,
  \frac{v_1^3-v_2^3}{(v_1^3+v_2^3)^3}\,
  \bigl[4](\frac{6\bar v_1^2\,(2\bar v_2^3-\bar v_1^3)}{(\bar v_1^3+\bar v_2^3)^4}\bigr[4])
\\
  &=
  -\frac{\Gamma^2}{6\uppi}\,
  \bigabs[4]{\frac{(v_1^3+v_2^3)^3}{v_1^3-v_2^3}}^2
  \frac{\bar v_1^2(v_1^3-v_2^3)(v_1^3+v_2^3)(2\bar v_2^3-\bar v_1^3)}{\abs{v_1^3+v_2^3}^8}\,
\\
  &=
  -\frac{\Gamma^2}{6\uppi}\,
  \frac{v_1^6-v_2^6}{\abs{v_1^6-v_2^6}^2}\,
  \bar v_1^2(2\bar v_2^3-\bar v_1^3)
\]
\vspace{-\abovedisplayskip}\vspace{-\belowdisplayskip}
\[
  \label{eq:v-u0-vector-field}
  \frac{dv_1}{dt}=
  -\frac{\upi\Gamma}{3\uppi}\,
  \frac{v_1^6-v_2^6}{\abs{v_1^6-v_2^6}^2}\,
  \bar v_1^2(2\bar v_2^3-\bar v_1^3)
\]
and $dv_2/dt$ is the same after exchange of $v_1$ and $v_2$.

\leaf{}{}
Using \eqref{eq:u-section} and \eqref{eq:u-to-v-dim3-reduction} to pull back the $\SE2$
momentum $\mu$ (see \eqref{eq:z-momentum})
\[
  &\mu=-\frac\Gamma6(3\abs{z_0}^2-\abs{z_1}^2-\abs{z_2}^2-\abs{z_3}^2),
\\
  &z_0=0,\qquad z_n=a\bigl[3](u_n+\frac{u_0}N\bigr[3])=a(u_n+u),\qquad \abs a=1,
\\
  &
  u=0,
  \qquad
  u_1=v_1+v_2,
  \qquad
  u_2=\uptheta v_1+\uptheta^2 v_2,
  \qquad
  u_3=\uptheta^2 v_1+\uptheta v_2,
  \qquad
  \uptheta=e^{2\uppi\upi/3},
\]
\vspace{-\belowdisplayskip}\vspace{-\abovedisplayskip}
\[
  \mu&=\frac\Gamma6(\abs{u_1}^2+\abs{u_2}^2+\abs{u_3}^2)
\\
   &=\frac\Gamma6(\abs{v_1+v_2}^2
   +\abs{v_1+\uptheta v_2}^2+\abs{\uptheta v_1+v_2}^2)
\\
  &=\frac\Gamma6\bigl(3\abs{v_1}^2+3\abs{v_2}^2
  +2\onm{Re}\bigl((1+\uptheta+\bar\uptheta)v_1\bar v_2\bigr)
\\
\label{eq:v-u0-momentum}
  &=\frac\Gamma2(\abs{v_1}^2+\abs{v_2}^2),
\]
and the corresponding vector field is
\[
  \frac{dv_n}{dt}
  =\frac{-1}\Gamma\times-2\upi\frac{\partial\mu}{\partial\bar v_n}
  =\frac{2\upi}\Gamma\times\frac\Gamma2
  =\upi v_n
  \label{eq:v-u0-infinitesimal-generator}
\]
ie $\mu$ is a momentum map for the diagonal action of $\SO2$.

\subnode{2-subnode-4-relative-equilibria}{Relative equilibria}
\leaf{}{}
Generally, for a dynamical system with phase space $\sset p$ and symmetry $\sset g$, an
evolution $c(t)$ of the form $g(t)p_e$, $g(0)=\bm1$, is by a one-parameter subgroup: for all
$s$, $t\mapsto c(t+s)$ is the evolution starting at $c(s)=g(s)p_e$ at time $t=0$, so by
equivariance $c(t+s)=g(s)(g(t)p_e)$, while $c(t+s)=g(t+s)p_e$, hence $g(t+s)=g(t)\,g(s)$. Thus
by Theorem \ref{thm:resolution-summary}, the relative equilibria in $P^\sing$ may be found from
the $\SO2$ relative equilibria in $P_1^\sing$.

\begin{theorem}
The rotational relative equilibria of the Hamiltonian system $P^0=\sset{z}$ are rotations,
translations, and $(z_1,z_2,z_3)$ permutations of (with corresponding equilibria on
$\partial P^1=\sset{v}$ the $\SO2$ orbits of)
\[
  &\upO_{\alpha_e}:\quad&&z=\alpha_e(0,1,e^{2\uppi\upi/3},e^{4\uppi\upi/3}),
  \quad
  v=(\alpha_e,0),
\\[2pt]
  &\upY_{\alpha_e}:\quad&&z=
  \frac{\alpha_e(1+r_1)}{2\sqrt{1+r_1^2}}\bigl[4](
  (-1,1,-2,-2)+\frac{\sqrt3\upi}2\frac{1-r_1}{1+r_1}\,(0,0,1,-1)\bigr[4]),
  \quad
  v=\frac{\alpha_e}{\sqrt{1+r_1^2}}\,(1,r_1),
\\[-2pt]
  &&&r_1=(1+\sqrt3-\sqrt[4]12)/2,
\]
and $u=\Gamma/3\uppi\alpha_e^2$, $\mu=\Gamma\alpha_e^2/2$, $v=0$, and $\nu=0$, for both
$\upO_{\alpha_e}$ and $\upY_{\alpha_e}$.
\end{theorem}

\leaf*{Proof}{}
The equations for the $\SO2$ relative equilibria on $P_1^\sing$ are
\[
  \label{eq:v-req-equations-1-1}
  &\frac{dv_1}{dt}=
  -\frac{i\Gamma}{3\uppi}\,
  \frac{v_1^6-v_2^6}{\abs{v_1^6-v_2^6}^2}\,
  \bar v_1^2(2\bar v_2^3-\bar v_1^3)
  =iu_ev_1,
\\
  \label{eq:v-req-equations-1-2}
  &\frac{dv_2}{dt}=
 -\frac{i\Gamma}{3\uppi}\,
  \frac{v_2^6-v_1^6}{\abs{v_2^6-v_1^6}^2}\,
  \bar v_2^2(2\bar v_1^3-\bar v_2^3)
  =iu_ev_2.
\]
where $u_e>0$.

\leaf{}{}
Substituting $v_2=0$ into \eqref{eq:v-req-equations-1-1}
\[
  \frac{\upi\Gamma}{3\uppi v_1}=iu_ev_1
  \quad\Leftrightarrow\quad
  u_e=\frac{\Gamma}{3\uppi v_1^2}.
\]
and \eqref{eq:v-req-equations-1-2} becomes $0=0$. Without loss of generality $v_1=\alpha_e$,
$\alpha_e>0$ (since $v_2=0$ and $\SO2$ acts diagonally on $(v_1,v_2)$), so
\[
  u_e=\frac{\Gamma}{3\uppi\alpha_e^2},
  \qquad
  \mu_e=\frac{\Gamma\alpha_e^2}2,
\]
and substituting into \eqref{eq:u-to-v-dim3-reduction} and using the standard
gauge \eqref{eq:u-section} obtains $\upO_{\alpha_e}$
\[
  \label{eq:z-req-O}
  z_0=0,
  \quad
  z_1=\alpha_e,
  \quad
  z_2=\alpha_e e^{2\uppi\upi/3},
  \quad
  z_3=\alpha_e e^{4\uppi\upi/3}.
\]
From \eqref{eq:se2-C-infinitesimal-generator}
\[
  &v_e=\frac{dz_0}{dt}
  =-\frac{i\Gamma}{6\uppi}\bigl[4](-\frac{z_1}{\abs{z_1}^2}
    -\frac{z_2}{\abs{z_2}^2}-\frac{z_3}{\abs{z_3}^2}\bigr[4])
  =0
\]
so the center of rotation is the origin.

\leaf{}{}
Without loss of generality assume $v_2\ne0$ and $v_1>0$. \eqref{eq:v-req-equations-1-1}
and \eqref{eq:v-req-equations-1-2} imply $v_2\,dv_1/dt-v_1\,dv_2/dt=0$, and removing nonzero
factors and conjugating is
\[
  \label{eq:v-req-equations-2}
  \bar v_2v_1^2(2v_2^3-v_1^3)=\bar v_1v_2^2(2v_1^3-v_2^3),
\]
which after substituting $v=v_1/v_2$ is
\[
  v^5-2\bar vv^3-2v^2+\bar v=0,
\]
and then substituting $v=re^{\upi\theta}$, $r>0$, and separating into real and imaginary
parts
\[
  \label{eq:v-req-equations-3}
  &r^4\cos6\theta-2r(r^2+1)\cos3\theta+1=0,
\\
  \label{eq:v-req-equations-4}
  &r^3\sin6\theta-2(r^2+1)\sin3\theta=0.
\]
If $\cos3\theta\ne 0$ and $\sin3\theta\ne0$ then the powers of $r$ may be iteratively eliminated
(eg form $\sin6\theta$ times the first minus $r\cos6\theta$ times the second) with result
\[
&r^3-\cos3\theta+r=0,
&&\qquad
r^3\cos3\theta-r^2-1=0,
\\
&r^2+r\cos3\theta-\cos^23\theta+1=0,
&&\qquad
r^2-r\cos3\theta+1=0,
\\
&2r-\cos3\theta=0,
&&\qquad
3r\cos^23\theta-2\cos^23\theta+2=0,
\]
so $\cos3\theta=0$ or $\sin3\theta=0$ because the last pair obtains the contradiction
$\cos^23\theta-4$. Substituting $\cos3\theta=0$ into \eqref{eq:v-req-equations-4} obtains the
contradiction $2(r^2+1)=0$, so $\sin3\theta=0$, $\cos3\theta=\pm1$,
and \eqref{eq:v-req-equations-3} becomes
\[
  r^4-2r(r^2+1)\cos3\theta+1=0.
\]
This is palindromic and the substitution $r+1/r=R$ obtains a quadratic equation
\[
 \label{eq:v-req-equations-5}
  R^2-2R\cos3\theta-2=0,
\qquad
  R=\cos3\theta\pm\sqrt3.
\]
Now $r+1/r=R$ is the quadratic $r^2-Rr+1=0$, which can have positive real roots only for $R>0$
(so $\cos3\theta=1$) and nonnegative discriminant $8(1\pm\sqrt3\cos3\theta)$ (so $R=1+\sqrt3)$),
after which the solutions are amoung the following six:
\[
  \theta=0,\;2\uppi,\;4\uppi;
  \qquad
  r_1\deq\frac12+\frac{\sqrt3}2-\frac{\sqrt{12}}2,
  \qquad 
  r_2\deq\frac12+\frac{\sqrt3}2+\frac{\sqrt{12}}2,
\]
Without loss of generality $\theta=0$, $v_1>0$, and $v_2=r_1v_1$, because
$r_1r_2=1$ and by \eqref{eq:S1-v-action} the permutation group $S_3$ acts on $v$ as
\[
  (1,2,3)\,v=e^{2\uppi\upi/3}v,
  \qquad
  (2,3)\,v=1/v,
\]
Using the standard gauge, the computations are
\[
  &z_0=0,
\\
  &z_1=v_1+v_2=(r_1+1)v_1,
\\
  &z_2=\theta v_1+\theta^2v_2=\theta v_1+\bar\theta v_2,
\\
  &\onm{Re}z_2=-\frac12 v_1-\frac12v_2=-\frac12(r_1+1)v_1,
\\
  &\onm{Im}z_2
  =\frac{\sqrt3}2v_1-\frac{\sqrt3}2v_2
  =\frac{\sqrt3(1-r_1)v_1}{2},
\\
  &z_2=\bigl[4](-\frac12(r_1+1)+\upi\frac{\sqrt3(1-r_1)}{2}\bigr[4])v_1,
\\
  &z_3=\theta^2 v_1+\theta v_2=\bar\theta v_1+\theta v_2=\bar z_2,
\\
  &v_e=\frac{dz_0}{dt}
  =-\frac{\upi\Gamma}{6\uppi}\bigl[4](-\frac{z_1}{\abs{z_1}^2}
    -\frac{z_2}{\abs{z_2}^2}-\frac{z_3}{\abs{z_3}^2}\bigr[4])
  =\frac{\upi\Gamma}{6\uppi}\bigl[4](\frac1{z_1}+\frac{2\onm{Re}z_2}{\abs{z_2}^2}\bigr[4]),
\\
  &\abs{z_2}^2=(\theta v_1+\bar\theta v_2)(\bar \theta v_1+\theta v_2)
  =v_1^2-v_1v_2+v_2^2
  =(1-r_1+r_1^2)v_1^2,
\\
  &v_e=\frac{\upi\Gamma}{6\uppi}\bigl[4](
  \frac1{(r_1+1)v_1}-\frac{r_1+1}{(r_1^2-r_1+1)v_1}\bigr[4])
  =\frac{-\upi\Gamma r_1}{2\uppi(r_1^3+1)v_1},
\\
  &\frac{2r_1^3-1}{r_1^6-1}=\frac1{r_1^2+1},
\\
  &u_e=\frac1{\upi v_1}\frac{dv_1}{dt}
  =\frac{(2r_1^3-1)\Gamma}{3\uppi(r_1^6-1)v_1^2}
  =\frac\Gamma{3\uppi(1+r_1^2)v_1^2},
\\ 
  &\frac{\upi v_e}{u_e}
  =\frac{\Gamma r_1}{2\uppi(r_1^3+1)v_1}
  \times\frac{\uppi(r_1+1)(r_1^3+1)v_1^2}{\Gamma r_1}
  =\frac12(r_1+1)v_1.
\]
After the scaling $v_1=\alpha_e/\sqrt{1+r_1^2}$, the $\SO2$ momentum becomes
\[
  \mu=\frac\Gamma2(\abs{v_1}^2+\abs{v_2}^2)
  =\frac\Gamma{2(1+r_1)^2}(\alpha_e^2+r_1^2\alpha_e^2)
  =\frac{\Gamma\alpha_e^2}2,
\]
with advantage that same $\alpha_e$ correspond same $\SO2$ symplectic reduced space as
$\upO_{\alpha_e}$, and then translating the center of rotation to the
origin obtains $\upY_{\alpha_e}$.~\nodeqed

\begin{figure}[t]
\begin{center}\includegraphics{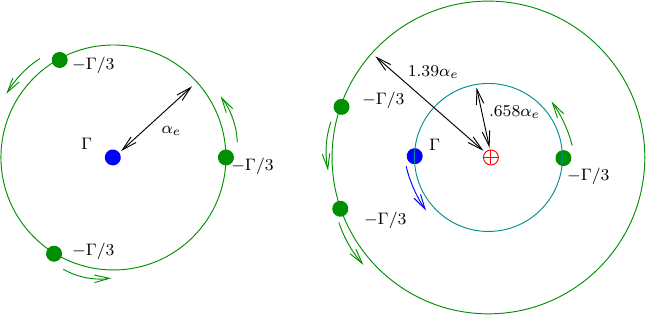}\end{center}
\caption{
  \label{fig:1-graphic-relative-equilibria} The relative equilibria $\upO_{\alpha_e}$ and
  $Y_{\alpha_e}$ as in the phase space $P=\sset{z}$. Both assemblages rotated rigidly as shown.}
\end{figure}

\leaf{}{}
The characteristic polynomial of the linearization of $\upO_{\alpha_e}$ is
$\lambda^2(\lambda^2+u_e^2)$ ie spectrally stable. The center of rotation of $\upY_{\alpha_e}$
is at
\[
  \frac{\upi v_e}{u_e}=\frac{3\uppi\alpha^2\upi}{\Gamma}
    \times\frac{-3^{1/4}\Gamma \upi}{6\uppi\alpha}
  =\frac{\alpha\sqrt[4]3}2
\]
ie half way between $z_0$ and $z_1$, and so $z_0$ and $z_1$ rotate diametrically opposite on an
inner circle while $z_2$ and $z_4$ rotate in the same circle on opposite sides of the line
between the first two. The ratio of the radius of the two circles is $\sqrt{1+2\sqrt3}=2.113$,
and the angle between the two outer vortices to the center or rotation is $37.62$ degrees, and
the characteristic polynomial is $\lambda^2(\lambda^2-(2\sqrt{3-\sqrt3}\,u)^2)$ ie spectrally
unstable. The heteroclinic orbits attached to $\upY_{\alpha_e}$ are exchanges of the outer
vortices.

\subnode{2-subnode-5-transdimensional-perturbation}{Transdimensional perturbation}

\begin{figure}[h]
\begin{center}\includegraphics[scale=1]{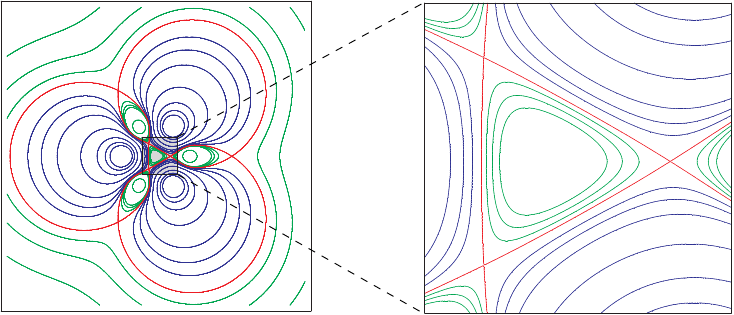}\end{center}
\caption{
  \label{fig:1-graphics-reduced-system}
  Energy levels of the $\SO2$ reduction of the Hamiltonian system~\eqref{eq:v-u0-hamiltonian}.}
\end{figure}

\leaf{}{} 
The $4$-vortex $\SE2$ symplectic reduced systems at zero translation momentum have dimension $2$
(the phase space has dimension $2\times 4=8$, while the momentum levels corresponding to the
zero dimensional coadjoint orbits in $\se2^{*\sing}$ and the $\SE2$ quotient subtract~$3$
each). By Thm.~\ref{thm:resolution-summary}, these (completely integrable) reduced systems are
the $\SO2$ reductions of $\bbC^2=\sset{(v_1,v_2)}$, with Hamiltonian~\eqref{eq:v-u0-hamiltonian}
and momentum \eqref{eq:v-u0-momentum}. Perturbative analyses to nearby $4$~dimensional
symplectic spaces are through \eqref{eq:v-hamiltonian} with small $u>0$; the dimension jump from
2 to 4 is spanned by the gauge group $\SO2$ (and its momentum).  The reduced spaces may be
realized by the substitution $v=v_2/v_1$ into~\eqref{eq:v-u0-hamiltonian} and then elimination
of $v_1$ using \eqref{eq:v-u0-momentum}
\[
  H=-\frac{\Gamma^2}{18\uppi}\,\onm{ln}\bigabs[4]{\bigl[4]
  (\frac\Gamma{2\mu}\bigr[4])^3\frac{(1+\abs v^2)^3(1-v^3)}{(1+v^3)^3}},
\]
so the energy level sets are those of the function of $v$ inside the logarithm
(Fig.~\ref{fig:1-graphics-reduced-system}).  $\upO_{\alpha_e}$ is the equilibrium in the center
and that is surrounded by three heteroclinic orbits between the three $\upY_{\alpha_e}$ (the other
$\upO_{\alpha_e}$ is at infinity), and there are six collision states. The periodic orbits near
$\upO_{\alpha_e}$ may be regarded as its internal dynamics.

\leaf{}{}
Perturbation of $\upO_{\alpha_e}$ can be accomplished by slice coordinates ie coordinates near
$v_1=\alpha_e$, $v_2=0$, which split into a part within momentum level sets and transverse to
the $\SO2$ orbit (and so coordinatizing the $\SO2$ reduced space), and another part for the
momentum and the group directions (and so coordinatizing
$T^*\SO2$)~\cite{PatrickGW-1995-1,PatrickGW-1999-1,RobertsM-WulffC-LambJSW-2002-1}. Here, these
are explicit, as follows: Seeking an submanifold transverse to the orbit and within the momentum
level set, is is natural to posit a graph of the form
\[
  (q,p)\mapsto\bigl(v_1(q,p),q-\upi p)\bigr),
\]
where $v_1(q,p)$ is real (the negative on $p$ is to remove and unwanted negative on the
coordinate symplectic form). Substituing this into the momentum and equating to the
$\upO_{\alpha_e}$ momentum
\[
  &\frac{\Gamma}2v_1^2+q^2+p^2=\frac{\Gamma}2\alpha_e^2,
  \qquad
  v_1=\sqrt{\alpha_e^2-q^2-p^2},
\]
where the positive root is so that the graph passes through the $\upO_{\alpha_e}$ at $q=0$,
$p=0$. To extend to other momenta, replace $\alpha_e$ with $2j+\alpha_e^2$
\[
  v_1=\sqrt{\alpha_e^2+2j-q^2-p^2}
\]
so that
\[
  \frac12(\abs{v_1}^2+\abs{v_2}^2)
  =\frac12(\alpha_e^2+2j-q^2-p^2)+\frac12(q^2+p^2)
  =j+\frac12\alpha_e^2,
\]
hence $j$ is the momentum relative the $\upO_{\alpha_e}$ momentum. Using the group action, the
slice becomes
\[
  v_1=\sqrt{\alpha_e^2+2j-q^2-p^2}\,\upe^{\upi\theta},
\qquad
  v_2=(q-\upi p)\,\upe^{\upi\theta},
\]
and an explicit local inverse at $\upO_{\alpha_e}$ is
\[
  q+\upi p=\frac{v_1\bar v_2}{\abs{v_1}},
  \quad
  j=\frac12(\abs{v_1}^2+\abs{v_2}^2-\alpha_e^2),
  \quad
  e^{\upi\theta}=\frac{v_1}{\abs{v_1}},
  \quad
  \abs{v_1-\alpha_e}<\alpha_e,
  \;
  -\uppi<\theta<\uppi.
\]
The symplectic form in the slice coordinates becomes $\Gamma(d\theta\wedge dj+dq\wedge dp)$,
because
\[
  &f\deq\sqrt{\alpha_e^2+2j-q^2-p^2},
\\
  &f\,df=f\times\frac1{2f}\times(2\,dj-2q\,dq-2p\,dp
  =dj-q\,dq-p\,dp,
\\
  &v_1=f\cos\theta+\upi f\sin\theta,
\\
  &v_2=(q\cos\theta+p\sin\theta)+\upi(q\sin\theta-p\cos\theta),
\\
  &d(f\cos\theta)\wedge d(f\sin\theta)
  =(df\cos\theta-f\sin\theta\,d\theta)\wedge(df\sin\theta+f\sin\theta\,d\theta)
  =f\,df\wedge d\theta,
\\
  &d(q\cos\theta+p\sin\theta)\wedge d(q\sin\theta-p\cos\theta)
  =-q\,d\theta\wedge dq-p\,d\theta\wedge dp-dq\wedge dp,
\\
  &-\Gamma d(f\cos\theta)\wedge d(f\sin\theta)
  -\Gamma d(q\cos\theta+p\sin\theta)\wedge d(q\sin\theta-p\cos\theta)
\\
  &\qquad\mbox{}=-\Gamma(dj-q\,dq-p\,dp)\wedge d\theta
  -\Gamma(-q\,d\theta\wedge dq-p\,d\theta\wedge dp-dq\wedge dp)
\\
  &\qquad\mbox{}=\Gamma(d\theta\wedge dj+dq\wedge dp).
\]

\leaf{}{}
In slice coordinates $\upO_{\alpha_e}$ corresponds to $q=p=j=0$ and the Hamiltonian can be
expanded near there using the substitution $v_1=fe^{\upi\theta}$, $v_2=\bar ze^{\upi\theta}$,
$z=q+\upi p$, and the rescaling $\epsilon q$, $\epsilon p$, and $\epsilon^2j$, with result
\[
  H&=
  \frac{\Gamma^2\ln\alpha_e}{3\uppi}
\\
  &\quad\mbox{}-\frac{\Gamma^2\Re(e^{\upi\theta}z)}{2\uppi\alpha_e^2}\,u
\\
  &\qquad\mbox{}+\frac{\Gamma^2(2j-\abs z^2)}{6\uppi\alpha_e^2}
  -\frac{3\Gamma^2\Re(z^2e^{2\upi\theta})}{4\uppi\alpha_e^4}\,u^2
\\
  &\qquad\quad\mbox{}+\frac{2\Gamma^2\Re(z^3)}{9\uppi\alpha_e^3}
  +\frac{\Gamma^2(2j-\abs z^2)\Re(e^{\upi\theta}z)}{2\uppi\alpha_e^4}\,u
\\
  &\qquad\qquad\mbox{}-\frac{\Gamma^2(2j-\abs z^2)^2}{12\uppi\alpha_e^4}
  +\frac{\Gamma^2\Re(z^4e^{\upi\theta})}{2\uppi\alpha_e^5}\,u
  +\frac{3\Gamma^2(2j-\abs z^2)\Re(z^2e^{2\upi\theta})}{2\uppi\alpha_e^6}\,u^2
\\
  &\qquad\qquad\quad\mbox{}+O(z^5,jz^3,j^2z,u^3),
  \label{eq:slice-hamiltonian-truncated-0}
\]
as has been obtained by Taylor expansion to order~$\epsilon^4$ followed by expansion
to~$u^2$. In the remainder term, $z^r$ denotes the set of homogeneous degree~$r$ polynomials in
$\Re z$ and~$\Im z$, and $O(S)$ denotes the ideal generated by~$S$ in the ring of smooth
functions. The $\SO2$ symmetry breaking at~$u=0$ corresponds to an absence of $\theta$ in the
expansion at~$u=0$. Terms involving powers of~$e^{\upi\theta}z$ have an
unexpected \nodetextdef{$\SO2$~low~order symmetry} of addition to~$\theta$ and simultaneous
reverse rotation of $z$ (this is not the $\SO2$ symmetry at $u=0$), arising because
$v_1v_2/(v_1^3+v_2^3)$ has degree $-1$ and at low order is multiplied by $z^1$. The expansion
has been carried to $\epsilon^4$ so as demonstrate low order symmetry breaking by the term
$z^4e^{2\upi\theta}$.

\begin{figure}[h]
\begin{center}\includegraphics[scale=1]{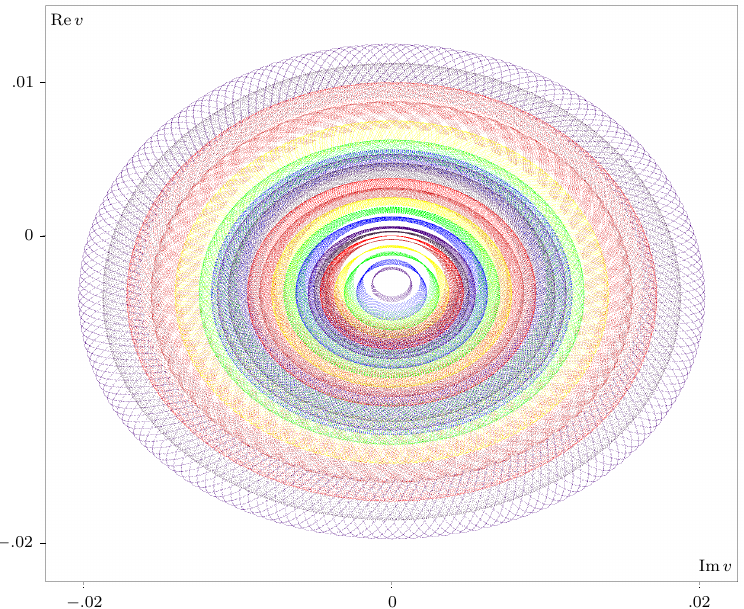}\end{center}
\caption{
  \label{fig:1-graphic-poincare-map}
  Poincare map near $\upO_{\alpha_e}$ corresponding to $\alpha_e=2$ and $u=.075$.}
\end{figure}

\leaf{}{}
At $u=0$, and truncating~\eqref{eq:slice-hamiltonian-truncated-0} at quadratic~$z$, the
corresponding differential equations are
\[
  \frac{d\theta}{dt}=\frac{\Gamma}{3\uppi\alpha_e^2}=\xi_e,
  \qquad
  \frac{dj}{dt}=0,
  \qquad
  \frac{dz}{dt}=-2\upi\,\frac{\partial H}{\partial\bar z}
  =\frac{\upi\Gamma}{3\uppi\alpha_e^2}z=i\xi_e z.
\]
Viewing $\upO_{\alpha_e}$ as an orbit, a Poincare section is $j=0$, $\theta=0$, and the Poincare
return map is the identity since the the periods of $\theta$ and $z$ are the same. Perturbation
of such a degenerate object is unlikely to yield useful information eg stability using the KAM
invariant curve theorem would required a twist
map~\cite{MeyerKR-HallGR-OffinD-2009-1,SiegelCL-MoserJK-1995-1}. Numerical computations of the Poincare map do
indicate stability but the usual invariant curves appear to be replaced with concentric
overlapping zones~(Fig.~\ref{fig:1-graphic-poincare-map}).

\leaf{}{}
The low order symmetry can be exploited to obtain a nondegenerate estimate of the Poincare
map. Removing constant terms, dividing the symplectic form and Hamiltonian by $\Gamma$, scaling
time as $t'\deq(\Gamma/3\uppi\alpha_e^2)t$, and setting $\epsilon=1$, and truncating, obtains
the $\SO2$-symmetric canonical Hamiltonian system
\[\label{eq:slice-hamiltonian-truncated-1}
  H&=j-\frac12\abs z^2-\frac{3u\Re(e^{\upi\theta}z)}{2}+\frac{2\Re(z^3)}{3\alpha_e}
  +O(u^2z^2,uz^3,z^4,juz,jz^2,u^3,j^2).
\]
Change to coordinates to $(\theta,k,Q,P)$ by 
\[
  k=j+\frac12\abs z^2,
\qquad
  w=Q+\upi P=e^{\upi\theta}z+\frac{3u}4.
\]
The function $k$ is the low order momentum, and, like $j$, scales as $\epsilon^2$, and these
are canonical coordinates because
\[
  &d\theta\wedge dk=d\theta\wedge dj+q\,d\theta\wedge dq+p\,d\theta\wedge dp,
\\
  &dQ\wedge dP=d(q\cos\theta-p\sin\theta+3u/4)\wedge(q\sin\theta+p\cos\theta)
  =dq\wedge dp+q\,dq\wedge d\theta+p\,dp\wedge d\theta,
\\
  &d\theta\wedge dk+dQ\wedge dP=d\theta\wedge dj+dq\wedge dp.
\]
The Hamiltonian~\eqref{eq:slice-hamiltonian-truncated-1} becomes (the remainder is
abbreviated in the first line and a constant is deleted)
\[
  H&=
  k-\bigabs[4]{w-\frac{3u}4}^2
  -\frac{3u}2\Re\bigl[4](w-\frac{3u}4\bigr[4])
  +\frac{2}{3\alpha_e}\Re\bigl[5](e^{-3\upi\theta}\bigl[3](w-\frac{3u}4\bigr[3])^3\bigr[5])
  +\mbox{h.o.t.}
\\
  &=k-\abs w^2
  -\frac{3u\Re(e^{-3\upi\theta}w^2)}{2\alpha_e}
  +O(ku^2,kuw,kw^2,u^3,u^2w,w^3,k^2),
  \label{eq:slice-hamiltonian-truncated-2}
\]
and the Hamiltonian vector field of~\eqref{eq:slice-hamiltonian-truncated-2} (the remainder
is supressed) is
\[
  &H=k-w\bar w-\frac{3u(e^{-3\upi\theta}w^2+e^{3\upi\theta}\bar w^2)}{4\alpha_e},
\\
  &
  \frac{d\theta}{dt}=\frac{\partial H}{\partial k}=1,
  \quad
  \frac{dk}{dt}=-\frac{\partial H}{\partial \theta}
    =-\frac{9\upi u\Re(e^{-3\upi\theta}w^2)}{2\alpha_e},
  \quad
  \frac{dw}{dt}
  =2\upi w+\frac{3iue^{3\upi\theta}\bar w}{\alpha_e}.
  \label{eq:slice-hamiltonian-vector-field-truncated-2}
\]
Use the initial condition $\theta(0)=0$, so that $\theta=t$, and
solving the last equation of~\eqref{eq:slice-hamiltonian-vector-field-truncated-2} (a routine Fourier transform exercise), obtains
\[\label{eq:poincare-map-approximate}
  w(t)=\bigl(A\sqrt{\omega^-}\,e^{-\upi\omega^+t}
  -\bar A\sqrt{\omega^+}\,e^{-\upi\omega^-t}\bigr)e^{2\upi t},
  \qquad
  \omega^{\pm}=\frac{1\pm\sqrt{1-36u^2/\alpha_e^2}}2,
\]
where $A$ is a complex constant of integration, and after substituting
$t=2\uppi$,~\eqref{eq:poincare-map-approximate} implies a linear map estimate with matrix
conjugate to an $O(u^2)$ rotation ie a matrix $PAP^{-1}$ where
\[
  P\deq\bmat{\sqrt{1-6u/\alpha_e}&0\\0&\sqrt{1+6u/\alpha_e}},
  \qquad
  A\deq\bmat{\cos2\uppi\omega^-&-\sin2\uppi\omega^-\\\sin2\uppi\omega^-&\cos2\uppi\omega^-}.
\]

\subnode{2-thread-2-conclusions}{Conclusions}
\leaf{}{} 
That $\upO_{\alpha_e}$ can exhibit a collective motion of a massive point particle, both on the
plane and the sphere, and that a Lagrangian model near the $\SE2$ group orbit may be used to
describe that, is observed in \cite{PatrickGW-2000-1}, and the symplectic reduced space at zero
translational momentum is computed in \cite{PatrickGW-2000-2}. The main contribution here is, by
resolution of the Poisson singularity, that the $2$-dimensional Hamiltonian dynamics at the
singularity can be glued smoothly as a boundary of the regular sector of $4$-dimensional ones,
at the cost of an $\SO2$ symmetry which describes a redundancy of states. Since perturbation to
nonzero translational momentum is necessarily from the singular sector to the regular sector,
the emergent mass is inextricably linked to an $\SO2$ symmetry breaking. Paradoxically, the
resolved $\upO_{\alpha_e}$ becomes a relative equilibrium (so a motion) of \emph{redundant
states} ie in that view the mass emerges from a motion that is \emph{not really there}.

\leaf{}{} 
Once the dimensions have been equalized, the questions become ones of ordinary Hamiltonian
perturbation theory and slice coodinates at $\upO_{\alpha_e}$ and the
expansion~\eqref{eq:slice-hamiltonian-truncated-0}. However, the usual stability by confinement
of KAM invariant curves requires robustly incomensurate frequencies and that fails in the
extreme because of a $1$-$1$ resonance in the associated Poincare return map. Thus the stability
problem has been illuiminated but remains open. Notably, the
expansion~\eqref{eq:slice-hamiltonian-truncated-0} cannot continue convergently for small
$\upO_{\alpha_e}$ radius because of the divisions by
$\alpha_e$. By~\eqref{eq:poincare-map-approximate}, the resonance gives rise to an emergent
frequency $\omega$ of second order in the translational momentum. The wavelength of this as
$\upO_{\alpha_e}$ translates is
\[
  \lambda=\mbox{period}\times \mbox{speed}
  =\frac{\mbox{momentum}}{\mbox{frequency}\times\mbox{mass}}
  =\frac{2\pi p}{m\omega}.
\]
As $\omega$ is proportional to the square of the momentum perturbation ie $\omega=kp^2$, the
wavelength is
\[
  \lambda=\frac{2\pi p}{m\,k p^2}
  =\frac{2\pi/mk}p,
\]
which is the same form as the de Broglie wavelength $\lambda=2\pi\hbar/p$.

\leaf{}{}
It should be mentioned that the identification of mass by the division (as observed in
simulation) of translational momentum by velocity is spurious because it involves divisions of
quantities obtained from the Lie algebra $\se2$ and its dual $\se2^*$. For a compact group this
has more substance since there is an invariant metric and the two spaces are naturally
identified. For the noncompact group $\SE2$ the identification of the dual is an arbitrary
theoretical input and only mass ratios between different $\upO_{\alpha_e}$ are valid
predictions.

\leaf{}{}
Finally, from a purely theoretical perspective, the regular sector of the resolution of the
$N$-vortex system is fully Poisson reduced, leaving a symplectic boundary with an addition
symmetry. The resulting \emph{Poisson manifold with boundary} may be regarded as a partial
reduction. In principle, the process may be applied to the boundary and iterated until no
symmetry remains, and thus viewed as a replacement for reduction itself. At the beginning of
modern Poisson geometry, from Alan Weinstein~\cite{WeinsteinA-1983-1},
\begin{quote}
\emph{The aim\,$\ldots$\,is to develop the theory of Poisson manifolds with an eye toward
these applications and also a new application\,---\,the study of singular limits of hamiltonian
systems.}
\end{quote}
While the $N$-vortex system is a coarse approximation of the full hydrodynamics, the resolution
of the $N$-vortex system derived here does seem to be consistent with this vision.

\begin{appendices}

\subnode{2-subnode-2-wirtinger-derivatives}{Wirtinger derivatives}
\label{subnode:wirtinger-derivatives}
\leaf{}{}
A complex valued function of two real variables $f(x,y)$ may be exchanged with a complex valued
function of two complex variables $f(z,\bar z)$ by the substitutions
\[
  z=x+\upi y,\quad\bar z=x-\upi y;
  \qquad
  x=\frac12(z+\bar z),\qquad y=\frac1{2\upi}(z-\bar z),
\]
and the \nodetextdef{Wirtinger derivatives} are by definition
\[
  \label{eq:wirtinger-derivatives}
  \frac{\pl f}{\pl z}
  \deq\frac12\bigl[5](\frac{\pl f}{\pl x}-\upi\frac{\pl f}{\pl y}\bigr[5]),
  \quad
  \frac{\pl f}{\pl\bar z}
  \deq\frac12\bigl[5](\frac{\pl f}{\pl x}+\upi\frac{\pl f}{\pl y}\bigr[5]),
\]
regarded as expressions in $z$ and $\bar z$ eg if $f(x,y)\deq xy$ then
$f(z,\bar z)=(z^2-\bar z^2)/4\upi$ and
\[
  \frac{\pl f}{\pl z}
  =\frac12(y-\upi x)
  =\frac1{2\upi}z,
  \qquad
  \frac{\pl f}{\pl\bar z}
  =\frac12(y+\upi x)
  =-\frac1{2\upi}\bar z.
\]
  If $f(x,y)=u(x,y)+\upi v(x,y)$ then
\[
  \label{eq:wirtinger-derivative-complex-derivative}
  &\frac{\pl f}{\pl z}
  =\frac12\bigl[4](\frac{\pl}{\pl x}-\upi\frac{\pl}{\pl y}\bigr[4])(u+\upi v)
  =\frac12\bigl[4](\frac{\pl u}{\pl x}+\frac{\pl v}{\pl y}\bigr[4])
  +\frac \upi2\bigl[4](\frac{\pl v}{\pl x}-\frac{\pl u}{\pl y}
  \bigr[4]),
\\
  &\frac{\pl f}{\pl\bar z}
  =\frac12\bigl[4](\frac{\pl}{\pl x}+\upi\frac{\pl}{\pl y}\bigr[4])(u+\upi v)
  =\frac12\bigl[4](\frac{\pl u}{\pl x}-\frac{\pl v}{\pl y}\bigr[4])
  +\frac \upi2\bigl[4](\frac{\pl v}{\pl x}+\frac{\pl u}{\pl y}
  \bigr[4]),
\]
so, if $f(x,y)$ is differentiable, then $\bar z$ or $z$ is missing after conversion to
$f(z,\bar z)$ if and only if $f(z)\deq f(\onm{Re}z,\onm{Im}z)$ is holomorphic or antiholomorphic
respectively, and then the Wirtinger derivatives are same as the usual complex analysis
derivatives $df/dz$ and $df/d\bar z$.  Wirtinger derivatives extend analogously to any number of
variables, and they satisfy the usual calculus rules, and also
from \eqref{eq:wirtinger-derivatives}
\[
  \frac{\pl f}{\pl x}
  =\frac{\pl f}{\pl z}+\frac{\pl f}{\pl \bar z},
  \qquad
  &\frac{\pl f}{\pl y}
  =\upi\bigl[4](\frac{\pl f}{\pl z}-\frac{\pl f}{\pl \bar z}\bigr[4]).
\]
This can be efficient in an application that lends itself to complex arithmetic eg
\[
  f(x,y)\deq\frac{\bar z}{z},
  \qquad
  \frac{\pl f}{\pl x}
  =\frac{\pl f}{\pl z}+\frac{\pl f}{\pl \bar z}
  =-\frac{\bar z}{z^2}+\frac 1z
  =\frac{2\upi\onm{Im}z}{z^2}
\]
as compared to computing $\pl f/\pl x$ from
\[
  f(x,y)=\frac{x^2-y^2}{x^2+y^2}-\frac{2\upi xy}{x^2+y^2}.
\]

\leaf{}{}
A formula may once be converted to refer to derivatives wrt $z$ and $\bar z$ and then a given
function of $z$ and $\bar z$ may not need to be converted to $x$ and $y$ eg the Poisson bracket
of $f(z)$ and $g(z)$ is
\[
  \{f,g\}=
  \frac{\pl f}{\pl x}\,\frac{\pl g}{\pl y}
  -\frac{\pl f}{\pl y}\,\frac{\pl g}{\pl x}
  &=\upi\bigl[4](\frac{\pl f}{\pl z}+\frac{\pl f}{\pl\bar z}\bigr[4])
  \bigl[4](\frac{\pl g}{\pl z}-\frac{\pl g}{\pl\bar z}\bigr[4])
  -f\leftrightarrow g
  =-2\upi\bigl[4](
  \frac{\pl f}{\pl z}\,\frac{\pl g}{\pl\bar z}
  -\frac{\pl f}{\pl\bar z}\,\frac{\pl g}{\pl z}
  \bigr[4]).
\]
For example, if $f=z\bar z=x^2+y^2$, $g=(z^2+\bar z^2)/2=x^2-y^2$ then
\[
  \{f,g\}
  &=-2\upi\bigl[5](
  \frac{\pl f}{\pl z}\,\frac{\pl g}{\pl\bar z}
  -\frac{\pl g}{\pl z}\,\frac{\pl f}{\pl\bar z}
  \bigr[5])
\\
  &=-2\upi\bigl((\bar z)(\bar z)-(z)(z)\bigr)
\\
  &=-2\upi\bigl((x-\upi y)(x-\upi y)-(x+\upi y)(x+\upi y)\bigr)
\\
  &=-2\upi(-4\upi xy)
\\
  &=-8xy,
\]
\vspace*{-\belowdisplayskip}\vspace*{-\abovedisplayskip}
\[
  \{f,g\}=\frac{\pl f}{\pl x}\,\frac{\pl g}{\pl y}
  -\frac{\pl f}{\pl y}\,\frac{\pl g}{\pl x}
  =(2x)(-2y)-(2y)(2x)
  =-8xy.
\]
If $f=u+\upi v$ is complex valued and $H$ is real then
\[
  \label{eq:wirtinger-vector-field}
  \frac{df}{dt}
  =\frac{du}{dt}+\upi\frac{dv}{dt}
  =\{u,H\}+\upi\{v,H\}
  =\{u+\upi v,H\}
  =\{f,H\},
  \qquad
  \frac{dz}{dt}=\{z,H\}=-2\upi\frac{\pl H}{\pl\bar z},
\]
and also
\[
  \label{eq:wirtinger-bracket-conjugate}
  \{f,g\}^{\mathord-}
  =2\upi
  \bigl[4](\frac{\pl f}{\pl z}\,\frac{\pl g}{\pl\bar z}
  -\frac{\pl g}{\pl z}\,\frac{\pl f}{\pl\bar z}\bigr[4])^{\mathord-}
  =2\upi
  \bigl[4](\frac{\pl\bar f}{\pl\bar z}\,\frac{\pl\bar g}{\pl z}
  -\frac{\pl\bar f}{\pl z}\,\frac{\pl\bar g}{\pl\bar z}\bigr[4])
  =\{\bar f,\bar g\}.
\]

\leaf{}{}
Given
\[
  f(w_1,\ldots,w_n,\bar w_1,\ldots,\bar w_N),
  \qquad
   g(w_1,\ldots,w_n,\bar w_1,\ldots,\bar w_N),
  \qquad
  w_1=w_1(z,\bar z),\ldots,w_n=w_n(z,\bar z),
\]
\vspace{-\belowdisplayskip}\vspace{-\abovedisplayskip}
\[
  \label{eq:wirtinger-bracket-transformation}
  \{f,g\}&=
  -2\upi\bigl[4](\frac{\pl f}{\pl z}\,\frac{\pl g}{\pl\bar z}
  -\frac{\pl f}{\pl\bar z}\,\frac{\pl g}{\pl z}\bigr[4])
\\
  &=-2\upi\sum_{i,j}
  \bigl[4](
  \frac{\pl f}{\pl w_i}\,\frac{\pl w_i}{\pl z}
  +\frac{\pl f}{\pl\bar w_i}\,\frac{\pl\bar w_i}{\pl z}
  \bigr[4])
  \bigl[4](
  \frac{\pl g}{\pl w_j}\,\frac{\pl w_j}{\pl\bar z}
  +\frac{\pl g}{\pl\bar w_j}\,\frac{\pl\bar w_j}{\pl\bar z}
  \bigr[4])
  -(f\leftrightarrow g)
\\
  &=-2\upi\sum_{i,j}\biggl(
  \frac{\pl f}{\pl w_i}\,\frac{\pl g}{\pl w_j}
  \,\frac{\pl w_i}{\pl z}\,\frac{\pl w_j}{\pl\bar z}
  +\frac{\pl f}{\pl w_i}\,\frac{\pl g}{\pl\bar w_j}
  \,\frac{\pl w_i}{\pl z}\,\frac{\pl\bar w_j}{\pl\bar z}
\\
  &\qquad\qquad\qquad\mbox{}+\frac{\pl f}{\pl\bar w_i}\,\frac{\pl g}{\pl w_j}
  \,\frac{\pl\bar w_i}{\pl z}\,\frac{\pl w_j}{\pl\bar z}
  +\frac{\pl f}{\pl\bar w_i}\,\frac{\pl g}{\pl\bar w_j}
  \,\frac{\pl\bar w_i}{\pl z}\,\frac{\pl\bar w_j}{\pl\bar z}
  \biggr)
  -(f\leftrightarrow g,i\leftrightarrow j)
\\
  &=\sum_{i,j}\bigl[4](
  \{w_i,w_j\}\frac{\pl f}{\pl w_i}\,\frac{\pl g}{\pl w_j}
  +\{w_i,\bar w_j\}\frac{\pl f}{\pl w_i}\,\frac{\pl g}{\pl\bar w_j}
  +\{\bar w_i,w_j\}\frac{\pl f}{\pl\bar w_i}\,\frac{\pl g}{\pl w_j}
  +\{\bar w_i,\bar w_j\}\frac{\pl f}{\pl\bar w_i}\,\frac{\pl g}{\pl\bar w_j}
  \bigr[4]),
\]
If the $w_i$ are holomorphic then $\{w_i,w_j\}=\{\bar w_i,\bar w_j\}=0$ and
\[
  \{f,g\}
  &=\sum_{i,j}\{w_i,\bar w_j\}\frac{\pl f}{\pl w_i}\,\frac{\pl g}{\pl\bar w_j}
  +\sum_{i,j}\{\bar w_i,w_j\}\frac{\pl f}{\pl\bar w_i}\,\frac{\pl g}{\pl w_j}
\\
  &=\sum_{i,j}\{w_i,\bar w_j\}\frac{\pl f}{\pl w_i}\,\frac{\pl g}{\pl\bar w_j}
  +\sum_{i,j}\{\bar w_j,w_i\}\frac{\pl f}{\pl\bar w_j}\,\frac{\pl g}{\pl w_i}
\\
  &=\sum_{i,j}
  \{w_i,\bar w_j\}\bigl[4](
  \frac{\pl f}{\pl w_i}\,\frac{\pl g}{\pl\bar w_j}
  -\frac{\pl f}{\pl\bar w_j}\,\frac{\pl g}{\pl w_i}
  \bigr[4]).
\]

\end{appendices}

\end{document}